\documentclass[twopage,11pt] {article}

\usepackage{psfig}

\newcommand\Lunit   {erg s$^{-1}$}
\newcommand\funit   {erg cm$^{-2}$ s$^{-1}$}

\begin{document}

\title{Crust cooling curves of accretion-heated neutron stars} 
\author{Rudy Wijnands}         

\maketitle

\pagestyle{myheadings} 
\thispagestyle{plain}         
\markboth{Rudy Wijnands}{Crust cooling curves of accretion-heated neutron stars} 
\setcounter{page}{1}

\noindent
{\bf Abstract:} We discuss the recent efforts to use a sub-class of
neutron-star X-ray transients (the {\it quasi-persistent} transients)
to probe the properties of neutron-star crusts and
cores. Quasi-persistent X-ray transients experience accretion episodes
lasting years to decades, instead of the usual weeks to months of
ordinary, short-duration transients. These prolonged accretion
episodes should significantly heat the crusts of the neutron stars in
these systems, bringing the crusts out of thermal equilibrium with
their neutron-star cores. When these systems are back in quiescence,
i.e. when no more accretion onto the neutron-star surfaces occurs,
then the crusts should thermally radiate in X-rays, cooling them down
until they are again in thermal equilibrium with the cores. In this
chapter we discuss the recent X-ray monitoring campaigns we performed
(using the X-ray satellites {\it Chandra} and {\it XMM-Newton}) to
study several quasi-persistent neutron-star X-ray transients in their
quiescent states. These observations gave us, for the first time, a
detailed look into the cooling curves of accretion heated neutron-star
crusts. In this chapter, we discuss how these crust cooling curves can
provide insight into the structure of neutron stars.

\section{Introduction \label{section:intro}}

In X-ray binary systems a neutron star or a black hole accretes
matter from a nearby companion star, either by Roche-lobe overflow or
by wind accretion. Due to conservation of angular momentum the
accreted matter does not directly fall onto the compact object, but
forms an accretion disk around it. In this accretion
disk, angular momentum is transported outward as the matter spirals
in.  A large amount of gravitational energy (up to
\mbox{$\sim$10$^{38}$ erg s$^{-1}$}) is released when the matter
approaches the compact object, heating the inner accretion disk to
very high temperatures (\mbox{$\sim$10$^7$ K}) and causing it to emit X-rays.
X-ray binaries can be divided into high-mass X-ray binaries and
low-mass X-ray binaries (LMXBs) after the mass of the companion
star. In the first group the mass of the companion star usually
exceeds ten solar masses and the accretion is driven by its strong
wind, while in LMXBs the companion's mass is below one solar mass and
the accretion is driven by Roche-lobe overflow. Systems which contain
a donor star with a mass between one and ten solar masses are rare,
due to the low efficiency of wind accretion and the instability of the
mass transfer through Roche-lobe overflow when the donor is more
massive than the receiving star. LMXBs can be further classified into
{\it persistent} and {\it transient} sources depending on their
long-term X-ray variability (see
\S~\ref{subsection:outburst_phase}). A sub-group of the transient
sources (the {\it quasi-persistent} neutron-star X-ray transients) has
recently led to advances in our understanding of the properties of the
cores and crusts of neutron stars and it is this research that is the
focus of this chapter.

\subsection{Neutron-star X-ray transients: outburst phase 
\label{subsection:outburst_phase}}

\begin{figure}[t]
\begin{center}
\begin{tabular}{c}
\psfig{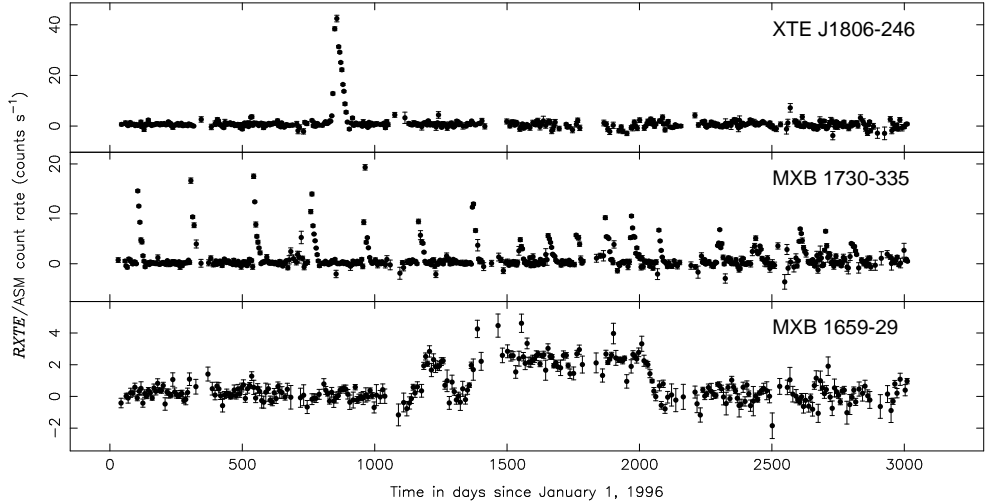}
\end{tabular}
\caption{
{\it RXTE}/ASM light curves (since January 1, 1996; i.e., since the
start of the {\it RXTE} mission) of the neutron-star X-ray transients
XTE J1806--246 (top), MXB 1730--335 (middle; this source is also known
as the Rapid Burster, located in the globular cluster Liller 1), and
MXB 1659--29 (bottom).  It is clear that a wide variety of outburst
durations and frequencies has been observed for different
transients. The ASM data points plotted represent four-day averages
for XTE J1806--246 and MXB 1730--335 , but seven-day averages for MXB
1659--29.
\label{fig:transients_lc} }
\end{center}
\end{figure}

The neutron-star X-ray transients form a special group among
neutron-star LMXBs.  They are usually very dim, with luminosities of
$10^{32-34}$~\Lunit, but occasionally they exhibit violent outbursts
during which their X-ray luminosity increases by several orders of
magnitude to $10^{36-38}$ \Lunit~(e.g., Chen {\em et al.}~1997).
These outbursts typically last for several weeks to months before the
systems turn off again.  For most of these sources only one outburst
has been observed, although several have shown multiple outbursts (see
Fig.~\ref{fig:transients_lc} for typical examples of light curves as
obtained with the all sky monitor [ASM] aboard the {\it Rossi X-ray
Timing Explorer} [{\it RXTE}] satellite).  These outbursts are very
likely the result of considerable increases in the mass accretion
rates onto the neutron stars in these systems, although the exact
physical processes behind these outbursts are still not well
understood (see, e.g., Lasota 2001 for a review of outburst
physics). Among the transients there is a special sub-class of sources
which do not turn off after a few weeks or months but which remain
active for many years (the 'quasi-persistent' transients). The best
examples of quasi-persistent transients harboring neutron stars are
EXO 0748--676 and GS 1826--238 (which are still active), and MXB
1659--29 and KS 1731--260 (which turned off recently).  In addition,
several neutron-star LMXBs which once were thought to be persistent
suddenly turned off (e.g., 4U 2129+47, X1732--304; XB 1905+000;
Pietsch {\em et al.}~1986; Guainazzi {\em et al.}~1999) and they
should also be considered quasi-persistent transients (throughout this
chapter when we refer to 'quasi-persistent transients' we mean {\it
quasi-persistent neutron-star X-ray transients} unless otherwise
noted\footnote{Note that also quasi-persistent {\it black-hole} X-ray
transients exist, such as GRS 1915+105 (active since May 1992;
Paciesas {\em et al.}~1994) and 4U 1630--47 (active since September
2002; Wijnands {\em et al.}~2002c), however, we do not discuss these
systems in this review.}).  During their outbursts, the neutron-star
transients (both the ordinary and the quasi-persistent transients) are
very similar to persistent neutron-star LMXBs with respect to their
X-ray properties and can be readily studied by the X-ray satellites
available. However, obtaining high quality X-ray data from the
transient systems during their quiescent state still remains a
challenge because of the much lower X-ray luminosities.

\subsection{Short duration neutron-star X-ray transients: quiescent phase}

Despite the low X-ray luminosities of neutron-star X-ray transient
during their quiescent state, they can still be detected with
sensitive imaging instruments. Although several intrinsically bright
and/or nearby systems were already detected with older generation
X-ray satellites (e.g., {\it EXOSAT}, {\it ROSAT}, {\it ASCA}, and
{\it BeppoSAX}; van Paradijs {\em et al.}~1987; Verbunt {\em et
al.}~1994; Garcia 1994; Asai {\em et al.}~1996, 1998; Campana {\em et
al.}~1998b, 2000; Garcia \& Callanan 1999; Stella {\em et al.}~2000),
the launch of the {\it Chandra} and {\it XMM-Newton} X-ray satellites
with their high sensitivity cameras meant a great leap forward in our
ability to detect quiescent systems and to obtain good X-ray spectra
(see, e.g., Daigne {\em et al.}~2002; in 't Zand {\em et al.}~2001;
Rutledge {\em et al.}~2001a, 2001b; Wijnands {\em et al.}~2001, 2002b,
2003; Campana {\em et al.}~2002; Jonker {\em et al.}~2003, 2004a).

\subsubsection{The observed X-ray properties in quiescence}

So far, the majority of quiescent neutron-star transients exhibit
X-ray spectra that are dominated by a soft ($<$$1$ keV) component
which can be accurately described by a thermal model such as a
black-body model or a modified black-body model like the neutron-star
atmosphere (NSA) models. In these NSA models it is assumed that the
depth from which the observed photons emerge from the neutron-star
atmosphere increases significantly with the energy of the photons, due
to the strong dependency of the opacities on energy.  Therefore, high
energy photons emerge from deeper and hotter layers than less
energetic photons. For a particular temperature, the emerging X-ray
spectrum is thus harder than the one that would result if the neutron
star would radiate as a pure black body. If indeed the neutron star
emits a NSA-like spectrum, then fitting that spectrum with a
black-body would overestimate the effective temperature (by a factor
of 2) and therefore underestimate the emitting area (often by an order
of magnitude; see, e.g., Zavlin {\em et al.}~1996 and references
therein).

The NSA models (those assuming a hydrogen atmosphere and a negligible
neutron-star magnetic field strength\footnote{The neutron stars in
X-ray transients are assumed to have magnetic field strengths of only
$10^8 - 10^9$ Gauss, which is sufficiently low not to affect the
spectra emerging from the neutron star. Thus the appropriate NSA
models are those which assume a zero magnetic field strength.}) have
recently dominated the spectral fits reported in the literature of
quiescent neutron-star X-ray transients. This is because NSA models
provide a clear physical explanation for the shape of the emitted
quiescent spectrum and they yield radii of the emitting area which are
consistent with the theoretically expected radii of neutron stars. In
contrast, black-body models typically give radii which are
significantly lower than those expected for neutron stars.  However,
it is important to stress that black-body models provide fits to the
data that are equally as satisfying as those of NSA models.  Thus,
currently, we cannot distinguish between these models observationally.

When fitting NSA models to the X-ray spectra of most quiescent
neutron-star transients, we observe that they have typically an
effective temperature (all effective temperatures in this chapter are
for an observer at infinity) of 0.1--0.2 keV and a bolometric
luminosity between $10^{32}$ and $10^{34}$ ergs s$^{-1}$. Several
systems have been found to exhibit an additional spectral component
which dominates the spectrum above a few keV and which can be
described by a simple power-law model (e.g., Asai {\em et al.}~1998;
Rutledge {\em et al.}~2001a, 2001b). This component can contribute up
to 50\% to the 0.5--10 keV quiescent flux of a particular system
(e.g., Rutledge {\em et al.}~2001b), although in other systems it
cannot be detected and at most 10\%--20\% of the 0.5--10 keV flux
might be due to such an additional hard spectral component (e.g.,
Wijnands {\em et al.}~2003).

Although most systems are dominated by the soft thermal component, two
systems are known {\it not} to follow this general trend. Instead,
they are dominated by the hard power-law component (contributing
$>$90\% to the 0.5--10 keV quiescent flux) and no thermal component
could be conclusively detected. Campana {\em et al.}~(2002) found that
the accretion-driven millisecond X-ray pulsar and X-ray transient SAX
J1808.4--3658 had a quiescent spectrum which was dominated by the hard
power-law component. Furthermore, its quiescent luminosity was
observed to be $5\times 10^{31}$ ergs s$^{-1}$, which makes it the
(intrinsically) faintest quiescent neutron-star transient currently
known. Very recently, Wijnands {\em et al.}~(2004b) found that, EXO
1745--248 in the globular cluster Terzan 5, is the second system with
a quiescent X-ray spectrum dominated by the hard power-law
component. Again the thermal component could not be detected and it
contributed at most 10\% to the quiescent 0.5--10 keV flux. Although
this resembles SAX J1808.4--3658, the 0.5--10 keV luminosity of EXO
1745--248 was a factor of 40 larger than that observed for SAX
J1808.4--3658. Currently, it is not understood why SAX J1808.4--3658
and EXO 1745--248 are different from the majority of quiescent
neutron-star X-ray transients.

\subsubsection{Theoretical models for quiescent neutron-star X-ray transients}

Several theoretical models have been developed to explain the low
quiescent X-ray luminosities and the X-ray spectra observed for
neutron-star X-ray transients. For example, the X-rays could be due to
the residual accretion of matter onto the neutron-star surface or down
to the magnetospheric boundary, or the pulsar emission mechanism
might be active (see, e.g., Stella {\em et al.}~1994; Zampieri et
al.~1995; Corbet 1996; Campana {\em et al.}~1998a; Menou {\em et
al.}~1999; Campana \& Stella 2000; Menou \& McClintock 2001). The
model currently most often used to explain the soft component is the
'cooling neutron star model'.

In this model (e.g., Campana {\em et al.}~1998a; Brown {\em et
al.}~1998) the radiation emitted below a few keV is thermal emission
originating from the neutron star surface.  Brown {\em et al.}~(1998)
argued that the neutron star core is heated by the nuclear reactions
occurring deep in the crust when the star is accreting. This heat is
released as thermal emission during quiescence. If the quiescent
emission is dominated by the thermal emission of the cooling neutron
star, then the quiescent luminosity should depend on the time averaged
(over $10^{4-5}$ years) accretion luminosity of the system (Campana
{\em et al.}~1998a; Brown {\em et al.}~1998).  Thus, the quiescent
luminosities observed can be directly compared with luminosities
predicted using estimations of the long term accretion history of the
sources.

The version of the cooling neutron star model presented by Brown et
al. (1998) was able to explain, at the time of its publication, the
luminosities of most of the systems then detected, although it was
found that the neutron-star transient Cen X-4 appeared to be less
luminous than this model predicted. This could be due to an
overestimation of the time-averaged accretion rate of Cen X-4 or due
to the existence of enhanced cooling processes in the core of the
neutron star (e.g., due to the direct Urca process, pion condensation,
or Cooper-pairing neutrino emission) instead of the standard core
cooling processes assumed by Brown {\em et al.}~(1998).  Since the
publication of the Brown {\em et al.}~(1998) paper, {\it Chandra} and
{\it XMM-Newton} have provided us with high quality data on about a
dozen quiescent neutron-star systems. We now know that Cen X-4 is not
the only system which is colder than expected on the basis of its
time-averaged accretion history and the standard cooling model (e.g.,
Campana {\em et al.}~2002; Nowak {\em et al.}~2002). It seems that two
groups of sources exist: those which can be explained by assuming
standard core cooling and those which require enhanced core cooling
processes.

Not all characteristics of the quiescent emission can be fully
explained by the cooling neutron star model (either using standard or
enhanced core cooling). For example, the neutron-star transients Aql
X-1, Cen X-4, and SAX J1748.9--2021 (located in the globular cluster
NGC 6440) have shown considerable variability in their quiescent
properties on time scales ranging from hundreds of seconds to years
(Rutledge {\em et al.}~2002a; Campana \& Stella 2003; Campana {\em et
al.}~2004; Cackett {\em et al.}~2004). This variability cannot easily
be explained by a cooling neutron star and extra ingredients need to
be added to the cooling model (e.g., Ushomirsky \& Rutledge 2001;
Brown {\em et al.}~2002) or alternative models must be used to explain
the quiescent properties (e.g., Campana \& Stella 2003; Campana {\em
et al.}~2004).  Furthermore, the power-law shaped spectral component
which dominates the quiescent spectra above a few keV in several
systems cannot be explained by the cooling models. The difficulty is
even more dire when attempting to explain the hard, power-law
component that dominates the quiescent X-ray spectra of SAX
J1808.4--3658 and EXO 1748--248. It is conceivable that this component
might be described by one or more of the alternative models mentioned
above (i.e., those which assume that the neutron star has a
non-negligible magnetic field strength). However, the observational
results on this component and our understanding of its nature are very
limited.

\section{Quasi-persistent neutron-star transients in quiescence}

Recently, we have demonstrated that quasi-persistent neutron-star X-ray
transients provide an excellent opportunity to improve our
understanding of the emission mechanisms at work in quiescent
neutron-star systems as well as elucidating the response of
neutron-stars to prolonged accretion periods (Wijnands {\em et
al.}~2001; Rutledge {\em et al.}~2002b). If the long durations of the
accretion episodes for these sources are typical (every outburst lasts
similarly long) and they are in quiescence for only decades (similar
to short-duration transients), then the cooling model predicts that
the cores of neutron stars in quasi-persistent systems should be
heated to considerably higher temperatures than those of the
short-duration transients, and consequently they should be more
luminous (factor of $>$10) in quiescence. This situation is further
complicated because the prolonged accretion has a considerable effect
on the crust (Rutledge {\em et al.}~2002b; for ordinary transients the
effect is much less). The crust will also be heated to very high
temperatures during the outburst and, depending on the crustal
relaxation time to return to thermal equilibrium with the core after
the outburst, the luminosity and temperature obtained for these
systems should reflect those of the crust and not of the core. The
exact crust relaxation time is unknown and estimates range from years
to decades. Monitoring observations of quasi-persistent systems in
their quiescent state following one of their prolonged accretion
events may allow for a determination of the structure of neutron stars
by comparing the observed cooling curves with those calculated (e.g.,
Rutledge {\em et al.}~2002b).  Recently, we have used two
quasi-persistent neutron-star transients (KS 1731--260 and MXB
1659--29) in quiescence to constrain neutron-star cooling curves and
the structure of neutron stars. Furthermore, at the time of writing
this chapter (April 2004), several other quasi-persistent neutron-star
X-ray transients have also been observed in quiescence. Below, we
discuss all quasi-persistent neutron-star systems observed, paying
particular attention to KS 1731--260 and MXB 1659--29.

\subsection{KS 1731--260}

\begin{figure}[t]
\begin{center}
\begin{tabular}{c}
\psfig{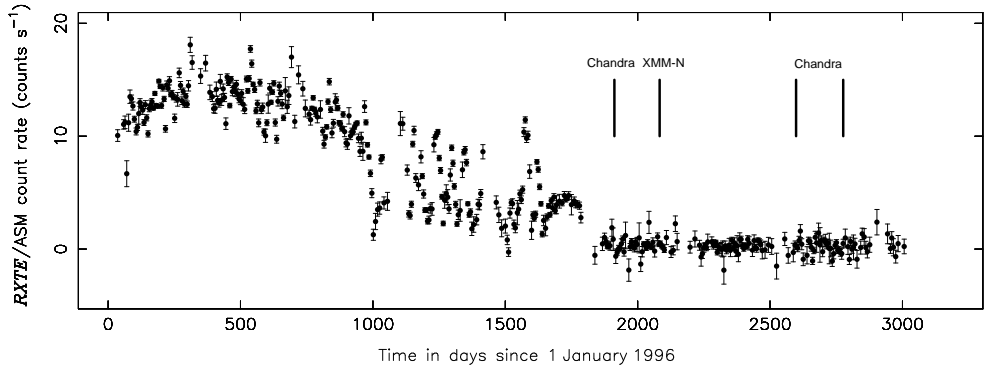}
\end{tabular}
\caption{
The {\it RXTE}/ASM light curve of KS 1731--260 since January 1,
1996. The beginning of the outburst was not observed since the source
first became active in 1989, more than 6 years before the launch of
{\it RXTE}.  The four solid lines show where our {\it Chandra} and
{\it XMM-Newton} observations were taken. The ASM data points are
5-day averages.
\label{fig:1731_asm} }
\end{center}
\end{figure}

\begin{figure}[t]
\begin{center}
\begin{tabular}{c}
\psfig{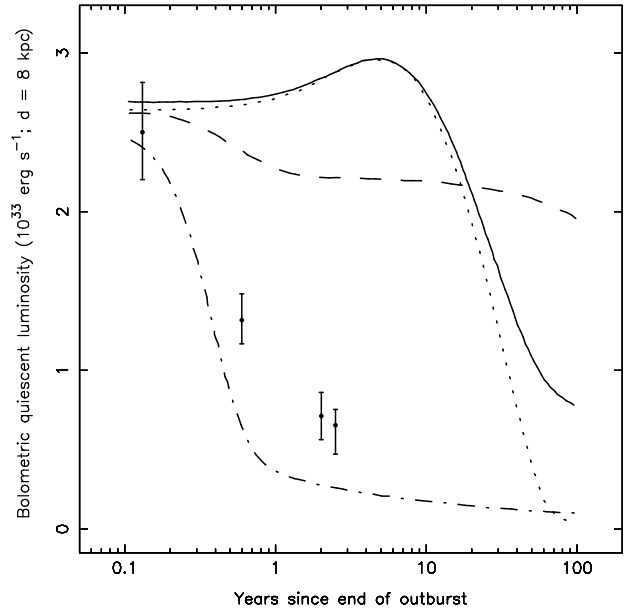}
\end{tabular}
\vspace{0.5cm}
\caption{
The cooling curves for KS 1731--260 calculated by Rutledge {\em et
al.}~(2002b). The solid line and dotted line assume low crustal
conductivity with standard or enhanced core cooling, respectively.
The dashed and dash-dotted lines correspond to large crustal
conductivity and standard or enhanced core cooling, respectively. The
luminosity measurements are given in the figure, although normalized
so that the first data point is consistent with the cooling
curves. The luminosity data points were obtained by fitting a NSA
model (the hydrogen NSA model for weakly magnetized neutron stars of
Zavlin {\em et al.}~1996) to the spectral data. The data points are
still preliminary and the full detailed analysis will be published by
Wijnands {\em et al.}~(2004c).
\label{fig:cooling}}
\end{center}
\end{figure}

KS 1731--260 was first discovered in August 1989 using the {\it
Mir}/Kvant instrument (Sunyaev 1989). The compact object in this new
transient was proven to be a neutron star based on the detection of
several type-I X-ray bursts\footnote{Type-I X-ray bursts are
thermonuclear explosions on the surface of accreting neutron
stars. Since a surface is needed to produce these bursts, observing
this phenomenon from a particular accreting system immediately rules
out a black hole in favor of a neutron star.} (Sunyaev 1989; Sunyaev
{\em et al.}~1990). Since its initial discovery, the source was
observed as a persistent source (Yamauchi \& Koyama 1990; Barret {\em
et al.}~1992, 1998; Smith {\em et al.}~1997; Wijnands \& van der Klis
1997; Muno {\em et al.}~2000; Narita {\em et al.}~2001) until the end
of 2000 or early 2001 when it suddenly turned off after having
actively accreted for over 12.5 years (Fig.~\ref{fig:1731_asm}). A
{\it Chandra} observation taken a few months after this transition
(Fig.~\ref{fig:1731_asm}) showed the source at a bolometric luminosity
of a few times $10^{33}$ \Lunit~(Wijnands {\em et al.}~2001,
2004c). Assuming that the long active period of KS 1731--260 is
typical for this source, the standard cooling model predicts that the
neutron-star core should be rather hot compared to that of ordinary,
short-duration transients, thus appearing as a very bright quiescent
system (as explained above). Furthermore, the long active period would
have heated the crust to high temperatures (Rutledge {\em et
al.}~2002b; much hotter than the core), which should make the system
even brighter. However, the low luminosity observed with {\it Chandra}
is very similar to what has been observed for the short-duration
systems in their quiescent states. Thus, the neutron-star crust and
core in KS 1731--260 were considerably colder than predicted by the
standard cooling model. Several scenarios could explain the unexpected
faintness of this source (Wijnands {\em et al.}~2001). First, it may
be that the recent accretion history of KS 1731--260 is atypical and
instead this source usually spends nearly a thousand years in
quiescence between outbursts (such long quiescent intervals are not
expected in the models proposed to explain X-ray outbursts;
e.g. Lasota 2001). Second, and more likely, it may be that the core of
the neutron star in this system undergoes enhanced cooling. This would
allow for the star to cool down quickly, making it very cold prior to
the last outburst episode. Then, during the outburst, the crust is
only heated to the observed temperature. Based on this first quiescent
observation of KS 1731--260, Rutledge {\em et al.}~(2002b) calculated
four cooling curves for the neutron star in this system, assuming
different microphysics for the core (standard vs. enhanced core
cooling) and the crust (a large vs. small heat conductivity; the heat
conductivity depends on the purity of the crust, with a large
conductivity when the crust is a pure crystal).

To follow the cooling of the neutron star (i.e., its crust) and to
test the calculated cooling curves of Rutledge {\em et al.}~(2002b),
KS 1731--260 was observed three more times, once with {\it XMM-Newton}
in 2001 (Wijnands {\em et al.}~2002b) and twice in 2003 using {\it
Chandra} (Wijnands {\em et al.}~2004c; Fig.~\ref{fig:1731_asm}). The
{\it XMM-Newton} observation found that the source had decreased in
luminosity within half a year by a factor of $\sim$2--3. This
indicated that the crust had cooled down further, which is consistent
with the observed decrease in the effective temperature (from $0.11$
to $0.09$ keV).  During a preliminary analysis (the details of the
analysis will be published by Wijnands {\em et al.}~2004c) of the
additional two {\it Chandra} observations, we found that the source
had decreased considerably (factor $\sim$8) in count rate compared to
the first observation (Fig.~\ref{fig:cooling}). We fitted the obtained
spectra using a hydrogen NSA model (for weakly magnetized neutron
stars; Zavlin {\em et al.}~1996) to determine how the bolometric
luminosity and its associated effective temperature decreased in
time. We found that the bolometric luminosity dropped from
$3\times10^{33}$
\Lunit~to $0.8\times10^{33}$ \Lunit~in $\sim$2.5 years
(Fig.~\ref{fig:cooling}) and the effective temperature decreased from
0.11 keV to 0.08 keV.  In the cooling model, this decrease in
luminosity means that the crust has cooled down significantly over
time.  When comparing the results with the cooling curves constructed
for KS 1731--260 (Rutledge {\em et al.}~2002b; Fig.~\ref{fig:cooling})
it is clear that none of these curves fit the data. However, the
curves have large uncertainties in their normalizations and in their
exact shapes due to the uncertainties in accretion history and
neutron-star properties (Rutledge {\em et al.}~2002b). The curve which
fits the data points best is the one which assumes a large crustal
heat conductivity and the presence of enhanced core cooling
processes. In addition, we see a possible leveling off of the
luminosity of KS 1731--260, indicating that the crust may soon come
into thermal equilibrium with the core and further observations of
this source in its quiescent state may yield the state of the core of
its neutron star. More detailed cooling curves must be calculated for
the neutron star in KS 1731--260 to determine if the cooling model can
fully explain the observations.

\subsection{MXB 1659--29}

MXB 1659--29 was discovered in 1976 by Lewin {\em et al.}~(1976)
during type-I X-ray bursts, which clearly established the compact
object in this system as a neutron star. The source was detected
several times between October 1976 and September 1978 with {\it SAS3}
and {\it HEAO} (e.g., Lewin {\em et al.}~1978; Share {\em et
al.}~1978; Griffiths {\em et al.}~1978; Cominsky {\em et al.}~1983;
Cominsky \& Wood 1984, 1989). At that time the source remained active
for $\sim$2.5 years (see Wijnands {\em et al.}~2003).  Pointed {\it
ROSAT} observations in the early 1990's failed to detect the source in
its quiescent state with an upper limit on the 0.5--10 keV flux of $(1
- 2) \times 10^{-14}$ \funit~(Verbunt 2001; Oosterbroek {\em et
al.}~2001; Wijnands 2002).

\begin{figure}[t]
\begin{center}
\begin{tabular}{c}
\psfig{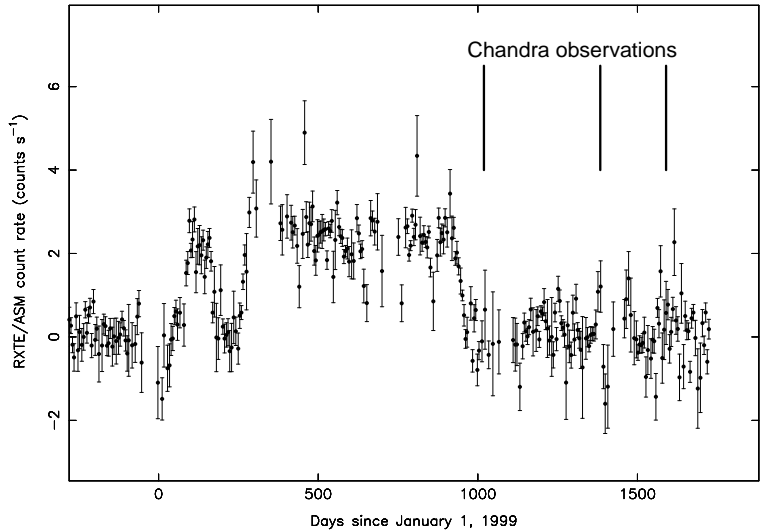}
\end{tabular}
\caption{
The {\it RXTE}/ASM light curve of MXB 1659--29 clearly showing the
1999--2001 outburst. The times of the {\it Chandra} observations are
indicated by the solid lines. All {\it Chandra} observations were
performed at times when the {\it RXTE}/ASM could not detect the
source. The ASM data points are 7-day averages.
\label{fig:1659_asm} }
\end{center}
\end{figure}

MXB 1659--29 remained dormant until April 1999, when in 't Zand et
al. (1999) reported it to be active again in observations obtained
with the {\it BeppoSAX} Wide Field Camera (see
Figs.~\ref{fig:transients_lc} and~\ref{fig:1659_asm}). The source was
observed on several occasions during this outburst using {\it
BeppoSAX}, {\it RXTE}, and {\it XMM-Newton} (e.g., Wachter {\em et
al.}~2000; Oosterbroek {\em et al.}~2001; Sidoli {\em et
al.}~2001). The source remained bright for almost 2.5 years before it
became quiescent again in September 2001 (Fig.~\ref{fig:1659_asm}).
Because of its long outburst durations, MXB 1659--29 can be classified
as a quasi-persistent source. It is interesting to note that the
outburst episode in 1999--2001 was similar in length to the outburst
episode in 1976--1978 which would suggest that such long outburst
durations are a common property of MXB 1659--29 (see Wijnands et
al. 2003 for a discussion).

\begin{figure}[t]
\begin{center}
\begin{tabular}{c}
\psfig{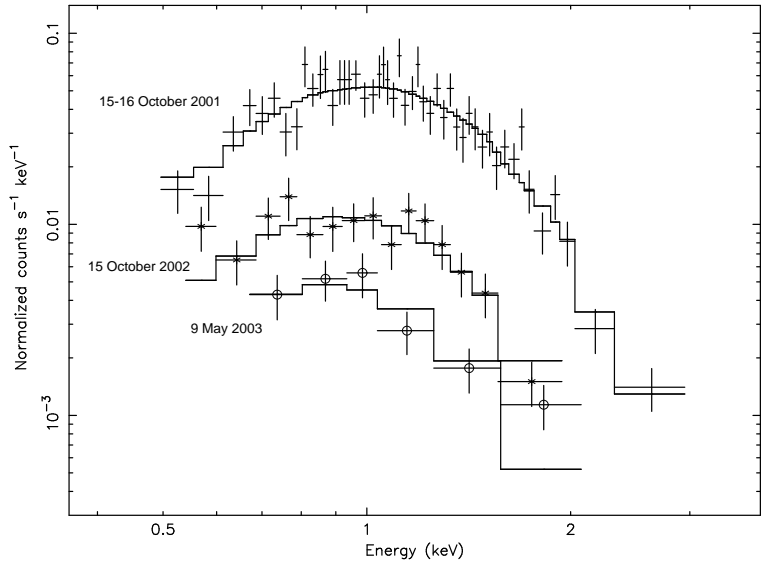}
\end{tabular}
\caption{The {\it Chandra} X-ray spectra obtained during the quiescent state of
MXB 1659--29. The top spectrum was obtained on October 15--16, 2001,
the middle spectrum (indicated by the crosses) was obtained on October
15, 2002, and the bottom spectrum (indicated by the open circles) was
obtained on May 9, 2003. The solid lines through the spectra indicate
the best fit neutron-star hydrogen atmosphere model (the one for
weakly magnetized neutron stars of Zavlin {\em et al.}~1996).
\label{fig:spectra} }
\end{center}
\end{figure}

\begin{figure}[t]
\begin{center}
\begin{tabular}{c}
\psfig{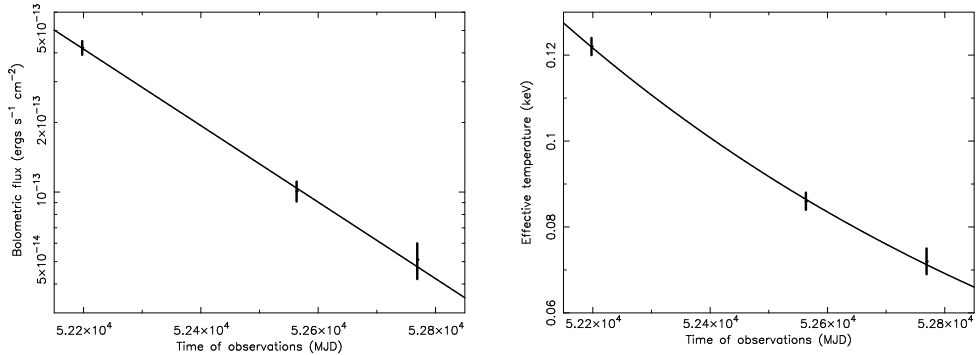}
\end{tabular}
\caption{
The bolometric flux (left) and effective temperature (right; for an
observer at infinity) of the neutron-star crust as a function of time
(as obtained with the neutron-star hydrogen atmosphere model for
weakly magnetized neutron stars of Zavlin {\em et al.}~1996). The
solid curves are the best fit exponential function through the data
points.  The bolometric fluxes are plotted on a logarithmic scale, but
for clarity, the effective temperatures are plotted on a linear scale.
\label{fig:decay} }
\end{center}
\end{figure}

A month after MXB 1659--29 turned off in 2001 we observed it using
{\it Chandra} (Fig.~\ref{fig:1659_asm}) and we found a thermal
spectrum with an effective temperature of 0.12 keV and a bolometric
luminosity of $5\times 10^{33}$ \Lunit~(Wijnands {\em et
al.}~2003). These properties are again very similar to those of
ordinary quiescent neutron-star systems despite the prolonged
accretion episode which would require the neutron-star crust, which
should be considerably out of thermal equilibrium with the core, to
dominate the X-ray emission observed.  Further monitoring observations
of MXB 1659--29 were performed in 2002 and 2003 (Wijnands {\em et
al.}~2004a; Fig.~\ref{fig:1659_asm}). Again the spectra of the source
were consistent with thermal radiation form the neutron-star surface
(Fig.~\ref{fig:spectra}). The bolometric flux decreased by a factor of
7--9 over the span of 1.5 years (Fig.~\ref{fig:decay}) and the rate of
decrease followed an exponential decay function.  Furthermore, the
effective temperature also decreased and the rate of decrease again
followed an exponential decay function. We found that the $e$-folding
time of the effective temperature curve was consistent with four times
that of the bolometric flux curve, as expected if the emission is
caused by a cooling black body for which the bolometric luminosity is
given by $L_{\rm bol}=4\pi\sigma R^{2}_{\infty} T^{\infty 4}_{\rm
eff}$ (with $T^{\infty}_{\rm eff}$ the effective temperature for an
observer at infinity and $L_{\rm bol}$ the bolometric luminosity of
the source).  This is consistent with the hypothesis that the
observations correspond to a cooling crust which was heated
considerably during the prolonged accretion event and which is still
out of thermal equilibrium with the core.

The cooling curves for the neutron star in KS 1731--260 (Rutledge {\em
et al.}~2002b) can be used as a starting point to investigate how the
results observed for MXB 1659--29 could be explained. Of those curves,
only the one which assumes a large crustal conductivity and the
presence of enhanced core cooling processes exhibits a large
luminosity decrease in the first two years after the end of the last
outburst, suggesting that the neutron star in MXB 1659--29 has similar
properties.  But detailed cooling curves for the neutron star in MXB
1659--29 need to be calculated to fully explore the impact of these
{\it Chandra} observations on our understanding of the structure of
neutron stars. The cooling curves calculated by Rutledge {\em et
al.}~(2002b) for KS 1731--260 only give us a hint of the behavior of
MXB 1659--29 because they depend on the long-term (\mbox{$>$10$^4$ years})
accretion history of the source. For KS 1731--260, this long-term
accretion behavior was quite unconstrained due to large uncertainties
in the averaged duration of the outbursts, the time-averaged accretion
rate during the outbursts, and the time the source spent in
quiescence. However, the accretion history of MXB 1659--29 over the
last three decades is much better constrained (Wijnands {\em et
al.}~2003), which will help to reduce the uncertainties in its
long-term averaged accretion history allowing for more detailed
cooling curves to be calculated for MXB 1659--29. This might help to
constrain the physics of the crust better for MXB 1659--29 than for KS
1731--260.

The 0.5--10 keV flux during the last {\it Chandra} observation
remained higher than the upper limit found with {\it ROSAT} (Verbunt
2001; Oosterbroek {\em et al.}~2001; Wijnands 2002), suggesting that
the crust will cool even further in quiescence and that the crust and
core have not yet reached thermal equilibrium. Further monitoring
observations are needed to follow the cooling curve of the crust to
determine the moment when the crust is thermally relaxed again.  When
this occurs, no significant further decrease of the quiescent
luminosity is expected and from this bottom level the state of the
core can be inferred. As of yet, we have found no evidence that the
flux and temperature are reaching a leveling-off value (Wijnands {\em
et al.}~2004a), associated with the temperature of the core, although
the limits we obtained are not very stringent.

\subsection{Alternative models \label{subsection:alternative}}

Our observations of KS 1731--260 and MXB 1659--29 demonstrated that
quasi-persistent sources can set tight constraints on the structure of
neutron stars. It is important to note that this assumes that the
observed properties are indeed due to a cooling crust and not due to
some other process.  Alternative models (e.g., residual accretion or
the onset of the radio pulsar mechanism; Campana {\em et al.}~1998a,
Campana \& Stella 2000) for the quiescent emission of KS 1731--260
have been proposed (e.g., Burderi {\em et al.}~2002).  Similarly,
Jonker {\em et al.}~(2004a) suggested that the difference in the
luminosity of MXB 1659--29 between the {\it ROSAT} non-detection and
the 2001 {\it Chandra} observation might be due to differences in
residual accretion rate onto the surface (residual accretion could
indeed produce soft spectra; e.g., Zampieri {\em et
al.}~1995). However, these models are far less predictive and thus
less verifiable than the cooling model. On the other hand, unlike the
cooling model, those alternative models predict that KS 1731--260 and
MXB 1659--29 should be similar to other systems if their neutron-star
parameters (i.e., spin rates and magnetic field properties) are
similar.  However, the most recent results on KS 1731--260 and MXB
1659--29 (Figs.~\ref{fig:cooling} and
\ref{fig:decay}) might pose extra problems for the alternative
models for the quiescent emission of both systems.  Although the
variations in bolometric luminosities for both sources can in
principle be explained by assuming variability in the accretion rate
or the amount of matter which interacts with the magnetic field, the
exponential decay curves we observe for the bolometric luminosities
and effective temperatures indicate that these alternative processes
must also decrease exponentially with a timescale of $\sim$1--2
years. Although this cannot be completely ruled out, we believe its is
unlikely since short-duration neutron-star transients have been
observed to reach their quiescent states on time-scales of tens to
several tens of days at the end of their outbursts (e.g., Aql X-1:
Campana {\em et al.}~1998b; RX J170930.2--263927; Jonker {\em et
al.}~2003\footnote{Recently, we became aware of the results of an
additional {\it Chandra} observation of RX J170930.2--263927 which was
performed $\sim$14 months after the last {\it Chandra} observation
reported in Jonker {\em et al.}~(2003). During this new observation
the source had a quiescent luminosity (P. Jonker private communication
2004; Jonker {\em et al.}~2004b) similar to that seen in the last
observation reported by Jonker {\em et al.}~(2003). This is further
confirmation that this source indeed returned to its lowest quiescent
luminosity within several tens of days after its outburst, in contrast
to the $>$2 years observed so far for KS 1731--260 and MXB 1659--29.}) 
and the variations in accretion rate tend to be more
stochastic. Moreover, if the neutron stars have significant magnetic
field strengths, then this might inhibit any material from reaching
the neutron-star surfaces when accreting at the inferred low accretion
rates. Despite this, alternative models should be considered when
interpreting the results obtained. Further monitoring observations of
both sources could rule out some models completely. In the cooling
model, large variations in the quiescent luminosities are not expected
anymore. If such large variations are observed then this would require
an alternative explanation for the quiescent emission. In particular,
large increases in luminosity might indeed be caused by a sudden surge
in the accretion rate.

\subsection{Other sources}

Several quasi-persistent neutron-star X-ray transients have been
observed in quiescence in addition to KS 1731--260 and MXB
1659--29. Unfortunately, those additional sources do not provide useful
constrains on the properties of the neutron-star crust and core. This
is because the observations were taken years after the end of the last
prolonged outburst episodes, the outburst histories (and thus the
accretion histories) of those sources are not well known, and the data
are of insufficient quality to constrain the X-ray spectra (and
sometimes even the bolometric luminosities). Below, I discuss briefly
these additional sources.

\subsubsection{EXO 0748--676}

This source was discovered with {\it EXOSAT} in February 1985 (Parmar
{\em et al.}~1986). Since then its has consistently been detected at
relatively high luminosities ($>10^{36}$ erg s$^{-1}$), and therefore,
this source can be regarded as a quasi-persistent transient which has
been active now for nearly 20 years.  Before its discovery, the source
had been serendipitously observed with {\it EINSTEIN} in May 1980 at a
quiescent luminosity of $\sim$$10^{34}$ erg s$^{-1}$ (Parmar {\em et
al.}~1986; Garcia \& Callanan 1999). Usually, this high quiescent
luminosity is explained as due to a relatively high level of residual
accretion during this particular observation (see, e.g., Garcia \&
Callanan 1999). However, if EXO 0748--676 is typically active {\it
and} quiescent for a few decades at a time, then from the Brown {\em
et al.}~(1998) model (and assuming standard core cooling and that the
long duration of the outburst episodes are typical for the source), we
would expect it to have a quiescent luminosity of $10^{34-35}$ erg
s$^{-1}$, consistent with what has been observed (see also Wijnands
2002).  So, even though KS 1731--260 does not behave as the simplest
version of the Brown et al. (1998) model predicts, EXO 0748--676 might
still do. {\it Chandra} or {\it XMM-Newton} observations are needed to
better constrain the quiescent bolometric luminosity and X-ray
spectrum of EXO 0748--676.  Unfortunately, the source has remained
active since it was originally discovered and no observations in
quiescence have yet been possible.

\subsubsection{X 1732--304 in the globular cluster Terzan 1}

In 1980, {\it Hakucho} detected a bursting X-ray source in the
direction of the globular cluster Terzan~1 (Makishima {\em et
al.}~1981; Inoue {\em et al.}~1981). Several years later, a steady
X-ray source was detected (X~1732--304) consistent with this globular
cluster and most likely corresponding to the bursting source (Skinner
{\em et al.}~1987; Parmar {\em et al.}~1989).  Since then, the source
has persistently been detected at 2--10 keV luminosities between a few
times $10^{35}$ erg s$^{-1}$ and $\sim$$10^{37}$ erg s$^{-1}$ (see
Figure 3 of Guainazzi {\em et al.}~1999 and references therein) until
1999, when Guainazzi {\em et al.}~(1999) reported an anomalous
low-state for X~1732--304 during a {\it BeppoSAX} observation. They
could only detect one dim source with a 2--10 keV luminosity of
$1.9\times10^{33}$ erg s$^{-1}$. This source luminosity and its X-ray
spectrum are both very similar to those observed for the quiescent
neutron star transients. This strongly indicates that X~1732--304
suddenly turned off and became quiescent after having accreted for
more than 12 years. This conclusion also holds if the {\it BeppoSAX}
source is not X~1732--304 but an unrelated source (which would likely
be also part of the globular cluster; Guainazzi {\em et
al.}~1999). The long active episode of X~1732--304 makes it a
quasi-persistent X-ray transient. Wijnands {\em et al.}~(2002a)
conducted a short {\it Chandra} observation of Terzan~1. Although they
detected one source with a 0.5--10 keV luminosity of \mbox{1-2$\times$
10$^{33}$}
\Lunit, they could not detect X~1732--304 conclusively, with a
bolometric luminosity upper limit of $1 - 2 \times 10^{33}$
\Lunit. Based on this luminosity upper limit and a detailed
examination of the accretion history of the source, Wijnands {\em et
al.}~(2002a) argued that, if this long outburst episode is typical for
X~1732--304 and the source stays dormant for only decades and not
centuries, then the neutron star in this source is colder than
expected from the standard cooling model requiring enhanced cooling
processes to be present in the core of the neutron star. The lack of a
detection of the source (and thus no known bolometric quiescent
luminosity) combined with the large uncertainties in its accretion
history inhibit strong conclusions for this system.

\subsubsection{4U 2129+47}

4U 2129+47 was discovered in 1972 (Giaconni {\em et al.}~1972; Gursky
{\em et al.}~1972) and, for a decade, it was seen by every X-ray
instrument which pointed at it (e.g., Markert {\em et al.}~1979;
Thorstensen {\em et al.}~ 1979; Ulmer {\em et al.}~1980; Brinkman {\em
et al.}~ 1980; Warwick {\em et al.}~1981; Amnuel {\em et al.}~1982;
Wood {\em et al.}~1984; Garcia 1994). In 1983 {\it EXOSAT} failed to
detect it (Pietsch {\em et al.}~1983, 1986) adding 4U 2129+47 to the
class of quasi-persistent X-ray transients.  The source has been
detected in quiescence with {\it ROSAT} (Garcia 1994; Garcia \&
Callanan 1998; Rutledge {\em et al.}~2000) and recently with {\it
Chandra} (Nowak {\em et al.}~2002). Its quiescent X-ray spectrum was
consistent with a thermal spectrum with a power-law component. If the
outburst and quiescent duration so far observed for 4U 2129+47 are
typical for this source, then the luminosity and temperature observed
are lower than expected from the standard cooling neutron star model
(Nowak {\em et al.}~2002; Wijnands 2002). However, the uncertainties
in the accretion history of the source hamper firm conclusions from
these results.  Note that all the quiescent observations were taken
many years after the end of the last prolonged outburst episode of 4U
2129+47, which likely means that, if the neutron star in 4U 2129+47
behaves similar to what we have observed for the neutron stars in KS
1731--260 and MXB 1659--29, during all these observations we measure
the state of the core and not of the crust since the crust has had
ample time to cool down and come into thermal equilibrium with the
core.

\subsection{XB 1905+000}

XB~1905+000 was first detected in 1974--1975 (Villa {\em et al.}~1976;
Seward {\em et al.}~1976) and was consistently seen to be active for a
decade afterwards (see Chevalier \& Ilovaisky 1990 for a historical
account until the mid-1980's). However, no reports of activity from XB
1905+000 are available after the {\it EXOSAT} observations of the
mid-1980's (e.g., Chevalier \& Ilovaisky 1990). Recently, we were
notified (P. Jonker 2003 private communication) that the source has
not been detected during a recent {\it Chandra}/HETG observation and
it is also not seen during the {\it ROSAT} and {\it ASCA} observations
presented in the public archives (A. Juett 2003, private
communication). For a distance of 8 kpc, the upper limit on the source
luminosity is $<$$8 \times 10^{32}$ ergs s$^{-1}$. These
non-detections suggest that the source may have been quiescent already
for over a decade, although the exact time when the source turned off
remains unclear.  If the neutron star in XB~1905+000 has similar
properties to those of the neutron stars in KS 1731--260 and MXB
1659--29, then it is expected that the quiescent X-ray luminosity of
XB~1905+000 should be dominated by the state of the core since the
crust should have already returned to thermal equilibrium with the
core. A further {\it Chandra} or {\it XMM-Newton} observation of this
source will likely detect the source in quiescence and this will give
us information about the state of the core of its neutron star.
However, because the accretion history of the source is not well
known, XB~1905+000 has limited usefulness in constraining the cooling
neutron star models.

\section{Conclusions}

Observations of quasi-persistent neutron-star X-ray transients (i.e.,
those of KS 1731--260 and MXB 1659--29) in quiescence have
demonstrated that this class of transients can be used to constrain the
structure of neutron stars.  However, in order to fully realize the
promise of these observations, more work still needs to be done,
especially calculating cooling curves specifically for each of the
neutron stars in the various quasi-persistent transients given that
these curves depend on the long-term accretion history of the source
which is quite different among systems.

Furthermore, additional quasi-persistent neutron-star transients
should be observed and monitored in their quiescent state preferably
within the first year after the end of their last accretion
episode. Such observations will allow us to determine whether the
behavior of KS 1731--260 and MXB 1659--29 is typical among
quasi-persistent neutron-star X-ray transients. Differences might be
expected not only because the accretion histories of the various
sources will differ significantly from each other but also because the
crusts (large heat conductivity vs. low heat conductivity) and cores
(standard vs. enhanced core cooling) of the neutron stars can behave
differently.

All the known persistent neutron-star LMXBs are promising candidates
should they ever turn off. As shown above, several sources which were
thought to be persistent have turned off suddenly, so it is possible
that an additional 'persistent' source might be found instead to be a
quasi-persistent X-ray transient. Unfortunately, the likelihood of any
one of them turning off is low since they have now been found to be
active for over 40 years (since the dawn of X-ray astronomy).  More
promising are those systems which have been seen to suddenly turn on
in the last 20 years and which have stayed on ever since (the most
promising candidates are EXO 0748--676, GS 1826--24, and XTE
J1759--220). It is likely that these systems may turn off in the
future and could then be used in a way similar to KS 1731--260 and MXB
1659--29 to constrain the properties of neutron stars.

\section*{Acknowledgments}

It is a pleasure to thank Peter Jonker for discussions about quiescent
neutron-star X-ray transients and for communicating the recent
behavior of RX J170930.2--263927 before publication. We also thank him
and Adrienne Juett for the information on XB~1905+000. We also thank
Luc\'\i a Mu\~noz Franco for help in preparing this chapter.

\section{References}

\parindent=0pt

Amnuel, P. R., Guseinov, O. Kh., \& Rakhamimov, Sh. Iu., 1982, {\it
Ap\&SS}, 82, 3

Asai, K., Dotani, T., Mitsuda, K., Hoshi, R., Vaughan, R., Tanaka, Y.,
\& Inoue, H., 1996, {\it PASJ}, 48, 257

Asai, K., Dotani, T., Hoshi, R., Tanaka, Y., Robinson, G. R., \&
Terada, K., 1998, {\it PASJ}, 50, 611

Barret, D., {\em et al.}, 1992, {\it ApJ}, 394, 615

Barret, D., Motch, C., \& Predehl, P., 1998, {\it A\&A}, 329, 965

Brinkman, A. C., {\em et al.}, 1980, {\it A\&A}, 81, 185

Brown, E. F., Bildsten, L., \& Rutledge, R. E., 1998, {\it ApJ}, 504,
L95

Brown, E. F., Bildsten, L., \& Chang, P., 2002, {\it ApJ}, 574, 920

Burderi, L. {\em et al.}, 2002, {\it ApJ}, 574, 930

Cackett, E., {\em et al.}, 2004, {\it ApJ}, submitted

Campana, S. \& Stella, L., 2000, {\it ApJ}, 541, 849

Campana, S. \& Stella, L., 2003, {\it ApJ}, 597, 474

Campana, S., Colpi, M., Mereghetti, S., Stella, L., \& Tavani,
M., 1998a, {\it A\&ARv}, 8, 279

Campana, S. {\em et al.}, 1998b, {\it ApJ}, 499, L65

Campana, S., Stella, L., Mereghetti, S., \& Cremonesi, D., 2000, {\it
A\&A}, 358, 583

Campana, S. {\em et al.}, 2002, {\it ApJ}, 575, L15

Campana, S., Israel, G. L., Stella, L., Gastaldello, F., \&
Mereghetti, S., 2004, {\it ApJ}, 601, 474

Chen, W., Shrader, C. R., \& Livio, M., 1997, {\it ApJ}, 491, 312

Chevalier, C. \& Ilovaisky, S. A., 1990, {\it A\&A}, 228, 115

Cominsky, L. R. \& Wood, K. S., 1984, {\it ApJ}, 283, 765

Cominsky, L. R. \& Wood, K. S., 1986, {\it ApJ}, 337, 485

Cominsky, L. R., Ossman, W., \& Lewin, W. H. G., 1983, {\it ApJ}, 270,
226

Corbet, R. H. D., 1996, {\it ApJ}, 457, L31

Daigne, F., Goldoni, P., Ferrando, P., Goldwurm, A., Decourchella, A.,
\& Warwick, R. S., 2002, {\it A\&A}, 386, 531

Garcia, M. R., 1994, {\it ApJ}, 435, 407

Garcia, M. R. \& Callanan, P. J., 1999, {\it AJ}, 118, 1390

Giaconni, R., Murray, S., Gursky, H., Kellog, E., Schreier, E., \&
Tananbaum, H., 1972, {\it ApJ}, 179, 281

Griffiths, R., {\em et al.}, 1978, {\it IAU Circ.}, 3190

Guainazzi, M., Parmar, A. N., \& Oosterbroek, T., 1999, {\it A\&A},
349, 819

Gursky, H. {\em et al.}, 1972, {\it ApJ}, 173, L99

Inoue, H., {\em et al.}, 1981, {\it ApJ}, 250, L71

in 't Zand, J., {\em et al.}, 1999, {\it IAU Circ.}, 7138

in 't Zand, J. J. M., van Kerkwijk, M. H., Pooley, D., Verbunt, F.,
Wijnands, R., \& Lewin, W. H. G., 2001, {\it ApJ}, 563, L41

Jonker, P. G., M\'endez, M., Nelemans, G., Wijnands, R., \& van der
Klis, M., 2003, {\it MNRAS}, 341, 823

Jonker, P. G., Wijnands, R., \& van der Klis, M., 2004a, {\it MNRAS},
349, 94

Jonker, P. G., {\em et al.}, 2004b, {\it ApJ} in preparation

Lasoat, J.-P., 2001, {\it NewAR}, 45, 449

Lewin, W. H. G., Hoffman, J. A., \& Doty, J., 1976, {\it IAU Circ.}, 2994

Lewin, W. H. G., {\em et al.}, 1978, {\it IAU Circ.}, 3190

Makishima, K. {\em et al.}, 1981, {\it ApJ}, 247, L23

Markert, T. H., {\em et al.}, 1979, {\it ApJS}, 39, 573

Menou, K. \& McClintock, J. E., 2001, {\it ApJ}, 557, 304

Menou, K., Esin, A. A., Narayan, R., Garcia, M. R., Lasota, J.-P., \&
McClintock, J. E., 1999, {\it MNRAS}, 305, 79

Muno, M. P., Fox, D. W., Morgan, E. H., \& Bildsten, L., 2000, {\it
ApJ}, 542, 1016

Narita, T., Grindaly, J. E., \& Barret, D., 2001, {\it ApJ}, 547, 420

Nowak, M. A., Heinz, S., \& Begelman, M. C., 2002, {\it ApJ}, 573, 778

Oosterbroek, T., Parmar, A. N., Sidoli, L., in 't Zand, J. J. M., \&
Heise, J., 2001, {\it A\&A} 376, 532

Paciesas, W. S., Deal, K. J., Harmon, B. A., Zhang, S. N., Wilson,
C. A., \& Fishman, G. J., 1996, {\it A\&AS}, 120, 205

Parmar, A. N., White, N. E., Giommi, P., \& Gottwald, M., 1986, {\it
ApJ}, 308, 199

Parmar, A. N., Stella, L., \& Giommi, P., 1989, {\it A\&A}, 222, 96

Pietsch, W., Steinle, H., \& Gottwald, M., 1983, {\it IAU Circ.}, 3887

Pietsch, W., Steinle, H., Gottwald, M., \& Graser, U., 1986, {\it
A\&A}, 157, 23

Rutledge, R. E., Bildsten, L., Brown, E. F., Pavlov, G. G., \& Zavlin,
V. E., 2000, {\it ApJ}, 529, 985

Rutledge, R. E., Bildsten, L., Brown, E. F., Pavlov, G. G., \&
Zavlin, V. E., 2001a, {\it ApJ}, 551, 921

Rutledge, R. E., Bildsten, L., Brown, E. F., Pavlov, G. G., \&
Zavlin, V. E., 2001b, {\it ApJ}, 559, 1054

Rutledge, R. E., Bildsten, L., Brown, E. F., Pavlov, G. G., \& Zavlin,
V. E., 2002a, {\it ApJ}, 577, 346

Rutledge, R. E., Bildsten, L., Brown, E. F., Pavlov, G. G.,
Zavlin, V. E., \& Ushomirsky, G., 2002b, {\it ApJ}, 580, 413

Stella, L., Campana, S., Colpi, M., Mereghetti, S., \& Tavani, M.,
1994, {\it ApJ}, 423, L47

Seward, F. D., Page, C. G., Turner, M. J. L., \& Pounds, K. A., 1976,
{\it MNRAS}, 175, 39

Share, G., {\em et al.}, 1978, {\it IAU Circ.}, 3190

Sidoli, L., Oosterbroek, T., Parmar, A. N., Lumb, D., \& Erd, C.,
2001, {\it A\&A}, 379, 540

Skinner, G. K., Willmore, A. P., Eyles, C. J., Bertram, D., \& Church,
M. J., 1987, {\it Nature}, 330, 544

Smith, D. A., Morgan, E. H., \& Bradt, H., 1997, {\it ApJ}, 479, L137

Stella, L., Campana, S., Mereghetti, S., Ricci, D., \& Israel, G. L.,
2000, {\it ApJ}, 537, L115

Sunyaev, R., 1989, {\it IAU Circ.}, 4839

Sunyaev, R. {\em et al.}, 1990, {\it SvAL}, 16, 55

Thorstensen, J., {\em et al.}, 1979, {\it ApJ}, 233, L57

Ulmer, M. P., {\em et al.}, 1980, {\it ApJ}, 235, L159

Ushomirsky, G. \& Rutledge, R. E., 2001, {\it MNRAS}, 325, 1157
	
van Paradijs, J., Verbunt, F., Shafer, R. A., \& Arnaud, K. A., 1987, {\it
A\&A}, 182, 47

Verbunt, F., 2001, {\it A\&A}, 368, 137

Verbunt, F., Belloni, T., Johnston, H. M., van der Klis, M., \& Lewin,
W. H. G., 1994, {\it A\&A}, 285, 903

Villa, G., {\em et al.}, 1976, {\it MNRAS}, 176, 609

Wachter, S., Smale, A. P., \& Bailyn, C., 2000, {\it ApJ}, 534, 367

Warwick, R. S., {\em et al.}, 1981, {\it MNRAS}, 197, 865

Wood, K. S., {\em et al.}, 1984, {\it ApJS}, 56, 507

Wijnands, R., 2002, in {\it The High Energy Universe at Sharp Focus}
proceedings of the 113th Meeting of the Astronomical Society of the
Pacific, p.235 (eds. E. M., Schlegel \& S. D. Vrtilek)

Wijnands, R. \& van der Klis, M., 1997, {\it ApJ}, 482, L65

Wijnands, R., Miller, J. M., Markwardt, G., Lewin, W. H. G.,
\& van der Klis, M., 2001, {\it ApJ}, 560, L159

Wijnands, R., Heinke, G. O., \& Grindlay, J. E., 2002a, {\it ApJ},
572, 1002

Wijnands, R., Guainazzi, M., van der Klis, M., \& M\'endez,
M., 2002b, {\it ApJ}, 573, L45

Wijnands, R., Remillard, R., \& Miller, J. M., 2002c, {\it ATEL}, 106

Wijnands, R., Nowak, M., Miller, J. M., Homan, J., Wachter, S., \&
Lewin, W. H. G., 2003, {\it ApJ}, 594, 952

Wijnands, R., Homan, J., Miller, J. M., \& Lewin, W. H. G.,
2004a, {\it ApJ}, 606, L61

Wijnands, R., {\em et al.}, 2004b, {\it ApJ}, in press
(astro-ph/0310144)

Wijnands, R., Homan, J., Miller, J. M., \& Lewin, W. H. G., 2004c,
{\it ApJ}, in preparation

Yamauchi, S. \& Koyama, K., 1990, {\it PASJ}, 42, L83

Zampieri, L., Turolla, R., Zane, S., \& Treves, A., 1995, {\it ApJ},
439, 849

Zavlin, V. E., Pavlov, G. G., \& Shibanov, Y. A., 1996, {\it A\&A}, 315, 141

\end{document}